# Asymmetric spin-wave dispersion due to Dzyaloshinskii-Moriya interaction in an ultrathin Pt/CoFeB film

Kai Di, Vanessa Li Zhang, Hock Siah Lim, Ser Choon Ng, Meng Hau Kuok[a]

*Department of Physics, National University of Singapore, Singapore 117551*

Xuepeng Qiu, and Hyunsoo Yang[b]

*Department of Electrical and Computer Engineering, National University of Singapore, Singapore 117576*

Employing Brillouin spectroscopy, strong interfacial Dzyaloshinskii-Moriya interactions have been observed in an ultrathin Pt/CoFeB film. Our micromagnetic simulations show that spin-wave nonreciprocity due to asymmetric surface pinning is insignificant for the 0.8nm-thick CoFeB film studied. The observed high asymmetry of the monotonic spin wave dispersion relation is thus ascribed to strong Dzyaloshinskii-Moriya interactions present at the Pt/CoFeB interface. Our findings should further enhance the significance of CoFeB as an important material for magnonic and spintronic applications.

[a] phykmh@nus.edu.sg.

[b] eleyang@nus.edu.sg.

Unlike the Heisenberg exchange interaction, the antisymmetric Dzyaloshinskii-Moriya interaction (DMI) favors canted neighboring spins.[1,2] DMI can arise at the interface between a ferromagnet and a nonmagnetic metal possessing strong spin-orbit coupling.[3] For ultrathin multilayer structures, such as Pt/Co, W/Fe, and Ir/Fe,[4-6] the effect of interfacial DMI on their spin dynamics can be significant, and thus an in-depth knowledge of this interaction is of importance. DMI is also responsible for the recently observed magnetic skyrmion crystals (SkXs),[6-9] which have been theoretically predicted much earlier.[10-13] The ultimate smallness and low-current-driven propagation of skyrmions make them promising for nanospintronics applications.[14,15] It has been shown that the SkX phase could be more stable in two-dimensional (2D) systems than in three-dimensional ones.[9,16]

Brillouin light scattering (BLS), the inelastic scattering of photons by excitations such as low-energy magnons, is a convenient and powerful tool for studying interfacial DMI in 2D systems, as it is a sensitive probe for detecting spin waves (SWs) in magnetic thin films.[17] In the BLS methodology, interfacial DMI is manifested as asymmetric dispersion relations and the nonreciprocal propagation of Damon-Eshbach (DE) SWs in multilayer film structures possessing broken inversion symmetry. While the propagation of DE SWs exhibits nonreciprocal surface localizations, it does not result in asymmetric dispersion relations for symmetrical magnetic films due to 2-fold rotational symmetry.[18] However, asymmetric multilayers could be subject to asymmetric surface pinning which would also result in the nonreciprocal dispersion relations of surface SWs. Indeed, such a pinning at opposite surfaces of a ferromagnetic film have been found to cause a significant difference in the frequencies of counter-propagating SWs.[19] An evaluation of the contribution from this effect to the asymmetric magnon dispersion is complicated, as it entails detailed knowledge of the pinning conditions of surface spins on various interfaces. However, this contribution can be made insignificant by making the magnetic film as thin as possible.

CoFeB is one of the most studied and widely used materials for spintronics devices and applications.[20-22] The highest tunneling magnetoresistance value at room temperature (RT) has been reported for a CoFeB magnetic tunneling junction.[23] Realization of a large DMI constant in CoFeB is highly anticipated due to its technological implication for chiral domain wall motion [24-26] and formation of helical spin spirals and magnetic SkXs.[6,14] We undertook a BLS study of the spin-wave dispersion relation of a Pt/CoFeB bilayer, where the CoFeB film is 0.8 nm thick. As the magnetic film is only about three atomic layers thick, the contribution from the above-mentioned asymmetric surface pinning to the observed asymmetric spin-wave frequency is negligible compared with that from interfacial DMI. Hence, a relatively accurate value of the DMI constant was obtained from the measured asymmetric magnon dispersion. We also provided a qualitative physical explanation of the observed asymmetric dispersion relation.

A MgO(2)/Pt(2)/$Co_{40}Fe_{40}B_{20}$(0.8)/$MgO$(2)/$SiO_2$(3) film [numbers in parentheses are the nominal thicknesses in nm, see inset of Fig. 1(a)] was deposited on thermally-oxidized Si substrate by high vacuum magnetron sputtering (base vacuum < $2\times10^{-9}$ Torr) at RT.[21] The sample will be referred to as Pt/CoFeB for short. Post-annealing of the film was performed at 240°C for 1 hour in high vacuum. The magnetic hysteresis loops of the sample (see Fig. 1), obtained separately from vibrating sample magnetometry (VSM) and magneto-optic Kerr effect (MOKE) measurements, indicate that the sample possesses partial perpendicular magnetic anisotropy (PMA) with an in-plane saturation field $H_{SAT} \approx 0.5$ T/$\mu_0$. The saturation magnetization was found to be $M_S = 1.6 \times 10^6$ A/m from the VSM measurement, which is within a similar range of reported values.[22,26,27] Assuming a second-order uniaxial anisotropy, the effective anisotropy field is estimated to be $H_U = 2K_U/\mu_0 M_S = H_{SAT} + M_S = 2.5$ T/$\mu_0$, corresponding to total surface anisotropy energy of about 1.6 mJ/m$^2$.

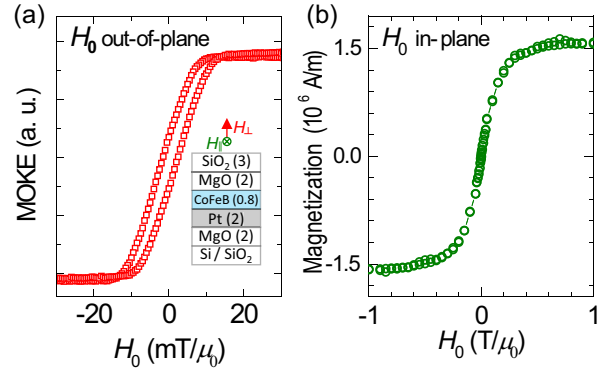

FIG. 1. (a) Out-of-plane and (b) in-plane magnetic hysteresis loops of the Pt/CoFeB film measured by MOKE and VSM, respectively. Inset: Schematics of the film structure and orientations of the applied magnetic field $H_0$.

As the top and bottom interfaces of the CoFeB film are different, the surface pinning may be asymmetric, resulting in the frequencies of counter-propagating SWs on opposite interfaces being different. By considering the case of highest asymmetric surface pinning, in which the surface anisotropy is assumed to be localized at one interface while the other is unpinned, we can estimate the upper limit of its contribution to the frequency asymmetry. Micromagnetic simulations using the OOMMF code were performed,[28,29] in which a $L_x \times L_y \times L_z$ = 4μm×4μm×0.8nm film was used. The film was discretized into 5nm×4μm×0.27nm mesh cells with three layers across the thickness to simulate the lattice discretization. The magnetic parameters are from our experiment measurements and fitting (see below). To calculate the highest frequency shift, we assume only the bottommost atomic layer possesses a uniaxial PMA with an anisotropy constant $K_U$ three times larger than the effective volume value from our measurement. Results reveal that the frequency difference $\Delta f$ due to asymmetric surface pinning is less than 0.02 GHz, even for the largest wavevector $k$ ($\approx$ 24 μm$^{-1}$) that can be attained in our BLS experiment. Importantly, this value is much smaller than the observed value of $\Delta f \approx$ 2 GHz (see below), and hence can be neglected. This result is expected, as the film is only some three-atom-layer thick, and hence, nonreciprocal localization of surface SWs is insignificant.

BLS spectra were recorded in the 180º-backscattering geometry at RT using a 6-pass Fabry-Perot interferometer.[17,30,31] Inelastic scattering from phonons has been excluded with the mutually perpendicular polarizations of the incident and scattered light. Mapping of the spin-wave dispersion relation was performed by changing the incidence angle $\theta$ of the $\lambda$ = 514.5 nm laser light [see Fig. 2(a)], and hence the magnon wavevector $k$.[32] With 30mW of laser power incident on the sample surface and an irradiated spot size of about 50 μm, heating of the sample was negligible. Although surface SWs propagating in $+x$ and $-x$ directions tend to localize on opposite surfaces of the CoFeB film, they were simultaneously detected, as the thickness of the CoFeB film is much smaller than the optical skin depth of metals. In the experiments, the equilibrium magnetization was oriented in-plane by applying a magnetic field $H_0 = 0.7$ T/$\mu_0$ in the DE geometry [see Fig. 2(a)]. Due to in-plane momentum conservation of the light scattering processing, SWs travelling in the $-x$ and $+x$ directions appear as peaks in the respective Stokes and anti-Stokes spectra.

Typical Brillouin spectra, recorded at various wavevectors and external fields $H_0$, are presented in Fig. 2(b). Pairs of Stokes and anti-Stokes peaks have asymmetric frequencies, with the frequency difference $\Delta f$ being more pronounced for larger spin-wave wavevectors. For a laser light incident angle of 60º (corresponding to wavevector $k$ = 21.2 μm$^{-1}$), the frequency difference $\Delta f$ is almost as large as 2 GHz. Clearly, based on the above numerical calculations, this large difference cannot be fully accounted for by asymmetric surface pinning, and hence it principally arises from interfacial DMI.

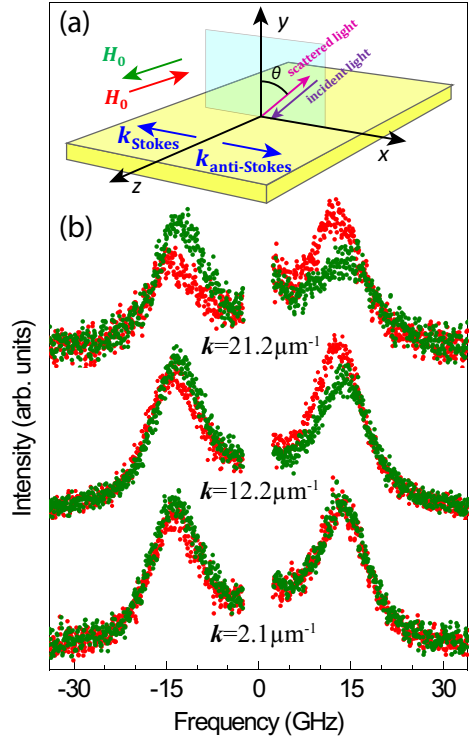

FIG. 2. (a) Schematics of Brillouin light scattering geometry, with scattering plane (in blue), and Cartesian coordinates. (b) Brillouin spectra of the Pt/CoFeB film measured at various spin-wave wavevectors. Green and red dots represent spectra recorded under applied field $H_0$ = 0.7 T/$\mu_0$ along +z and –z directions, respectively.

The measured spin-wave dispersion relation is shown in Fig. 3(a). Unlike the typical V-shaped dispersion curves of surface SWs in thicker magnetic thin films, the measured dispersion curve of the Pt/CoFeB sample is monotonic and basically linear for the range of $k$ studied. Solving the linearized Landau-Lifshitz equation which includes an effective DMI field term,[17,33,34] yields the following spin-wave dispersion relation

$$f = \frac{\mu_0 \gamma}{2\pi}\sqrt{\left[H_0 + Jk^2 + \xi(kL)M_S\right]\left[H_0 - H_U + Jk^2 + M_S - \xi(kL)M_S\right]} - \frac{\gamma}{\pi M_S}Dk, \quad (1)$$

where $\gamma$ is the gyromagnetic ratio, $J = 2A/(\mu_0 M_S)$ with $A$ being the exchange constant, $L$ (= 0.8 nm) the thickness of the CoFeB film, $\xi(x) = 1 - \left(1 - e^{-|x|}\right)/|x|$, and $D$ the DMI constant. It is

6-12

noteworthy that if $D = 0$, Eq (1) predicts that the magnon propagation will be almost dispersionless [see Fig. 3(a)], which is a consequence of the ultra-thinness of the film studied. The DMI-induced second term on the right hand side of Eq. (1) causes the dispersion curve to become an inclined straight line. This is indicative of the presence of very strong spin-wave nonreciprocity as the detected SWs possess a near-constant group velocity, irrespective of the sign of their wavevectors $k$. Interestingly, due to the DMI-induced linear term, the $+k$ SWs exhibit negative group velocities. The $D = 0$ dispersion curve of Fig. 3(a) shows that for small positive wavevectors, the SWs possess a negative group velocity, unlike typical DE-type SWs. It can easily be seen from Eq. (1) that this is a consequence of the PMA (and hence $M_S < H_U$) in our film.

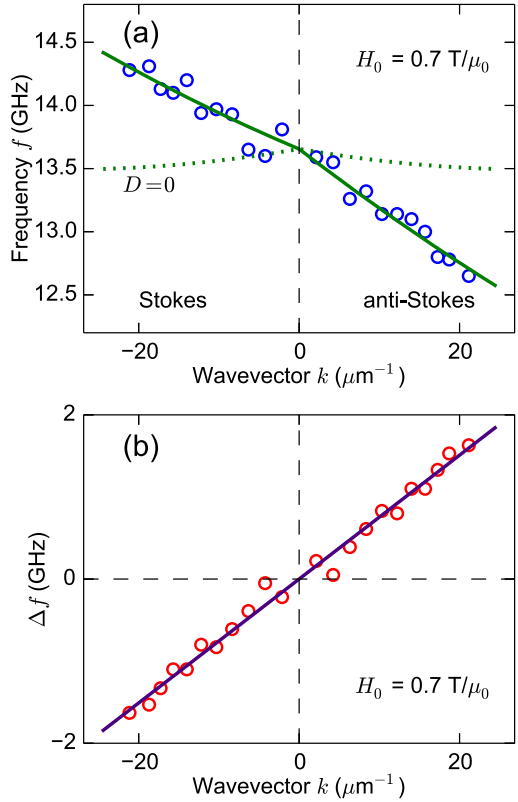

FIG. 3. (a) Spin-wave dispersion relation of the Pt/CoFeB film measured under an in-plane applied field $H_0 = 0.7$ T$/\mu_0$ along the $-z$ direction. (b) Frequency difference of counter-

propagating SWs as a function of wavevector. Open circles denote measured data. The symmetrical dotted lines in (a) represent the dispersion curve calculated for zero DMI, while solid lines in (a) and (b) represent respective fits to experimental data.

From Eq. (1), the difference in the frequencies of counter-propagating SWs is

$$\Delta f(k) = f(-k) - f(k) = \frac{2\gamma}{\pi M_S} Dk, \qquad (2)$$

which is linear in $D$ and $k$. The experimental variation of $\Delta f$ with wavevector $k$ is presented in Fig. 3(b). A linear fit to the experimental data, based on $\gamma$ = 190 GHz/T and the VSM-measured value of $M_S$ = 1.6×10$^6$ A/m, yields a DMI constant value of $D$ = 1.0±0.1 mJ/m$^2$, where the error is estimated from Eq. (2) taking into account the experimental uncertainties of $\Delta f$, $\gamma$, $M_S$, and $k$. This value represents the effective bulk DMI constant, i.e. the interfacial DMI constant averaged over the film thickness.[35] This effective bulk DMI constant is relatively large compared with those of other similar systems, such as Pt/NiFe ($D$ = 0.1 – 0.6 mJ/m$^2$) and Hf/CoFeB ($D$ = 0.5 mJ/m$^2$).[36,37] The $\gamma$ value [190×10$^9$ rad/(s·T)] used for our CoFeB film is reasonable as it is close to the respective typical values of 194 and 185 ×10$^9$ rad/(s·T) for Co and Fe.[38] The fitted exchange constant $A$ is about 5 pJ/m, which is smaller than typical values for CoFeB. The estimated 95% confidence interval of $A$ from the fitting is also very large (> 100%), meaning that an accurate value cannot be extracted from our data. However, from Eq. (2) it should be noted that $D$ is independent of $A$.

We next briefly discuss the underlying physics behind the DMI-induced nonreciprocal frequency shift. For simplicity, we assume that the dynamic magnetization of the SWs is circularly polarized and can be expressed as $\boldsymbol{m} = \boldsymbol{m}_x + \boldsymbol{m}_y = \boldsymbol{e}_x \sin(\omega t - kx) + \boldsymbol{e}_y \cos(\omega t - kx)$, where $\boldsymbol{e}_x$ and $\boldsymbol{e}_y$ are unit vectors along the +$x$ and +$y$ directions, respectively. The effective field arising from the DMI is[17,33]

$$h_{\text{DMI}} = -D^*\left(e_z \times \frac{\partial m}{\partial x}\right) = D^* k m, \qquad (3)$$

where $D^* = 2D/\mu_0 M_S$. Equation (3) shows that the dynamic effective DMI field is proportional to the wavevector $k$ and is either parallel or antiparallel to $m$. As Fig. 4 shows, the magnetization vector $M$ precesses under the influence of the total effective field $H_{\text{eff}} + h_{\text{DMI}}$ according to the Landau-Lifshitz equation $dM/dt = -\mu_0 \gamma M \times (H_{\text{eff}} + h_{\text{DMI}})$, where $\mu_0 \gamma > 0$, and the damping term has been neglected. Therefore, the magnitude of $M \times (H_{\text{eff}} + h_{\text{DMI}})$ determines the frequency of the precession. For SWs propagating along the $+x$ direction [$k > 0$, see Fig. 4(a)], $-M \times h_{\text{DMI}}$ is antiparallel to $-M \times H_{\text{eff}}$, and hence the precession frequency will be lower than that for zero DMI. In contrast, for SWs travelling in the $-x$ direction [$k < 0$, see Fig. 4(b)], $-M \times h_{\text{DMI}}$ is parallel to $-M \times H_{\text{eff}}$, resulting in a faster precession and hence higher frequencies. Furthermore, because $-M \times h_{\text{DMI}}$ is proportional to $k$, the above arguments also account for the experimental observation that $\Delta f$ is larger for larger wavevectors.

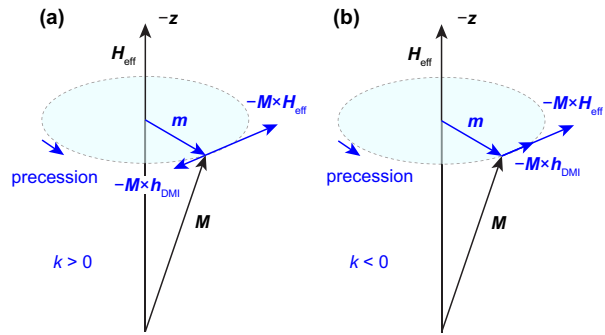

FIG. 4. Schematics of the precession of the magnetization $M$ under the total effective field $H_{\text{eff}} + h_{\text{DMI}}$ for (a) $k > 0$ and (b) $k < 0$. All the vectors in the $xy$-plane are labelled blue.

In summary, the spin-wave dispersion relation of a Pt/CoFeB film has been measured by Brillouin spectroscopy. The observed high asymmetry of the monotonic dispersion curve is attributed to the presence of interfacial DMI at the Pt/CoFeB interface. The contribution from asymmetric surface pinning to the observed dispersion asymmetry is made insignificant by choosing an ultrathin (0.8 nm) magnetic layer. Indeed, our calculation shows that this contribution is too small to fully account for the measured large difference in the frequencies of counter-propagating SWs. The DMI constant of the 0.8nm-thick CoFeB film was measured to be $D = 1.0$ mJ/m$^2$. Our study would be of use for understanding the interfacial DMI at Pt/ferromagnet interfaces and for DMI-related applications, such as nonvolatile storage and information processing based on chiral domain walls and skyrmions.

This project was funded by the Ministry of Education, Singapore under Academic Research Fund Grant No. R144-000-340-112 and the National Research Foundation, Prime Minister's Office, Singapore under its Competitive Research Programme (CRP Award No. NRF-CRP12-2013-01).